\PassOptionsToPackage{bookmarks=false}{hyperref}

\makeatletter
\def\input@path{{./figures/}}
\makeatother

\documentclass[format=sigconf,table]{acmart}

\usepackage[english]{babel}
\usepackage[utf8x]{inputenc}
\usepackage[T1]{fontenc}

\usepackage{multirow}

\usepackage{xspace}
\usepackage{amsmath}
\usepackage{graphicx}
\usepackage[colorinlistoftodos]{todonotes}
\usepackage{csquotes}

\usepackage{graphicx}
\usepackage{subcaption}

\usepackage{booktabs}
\usepackage[ruled,vlined,linesnumbered]{algorithm2e}
\let\oldnl\nl
\newcommand{\nonl}{\renewcommand{\nl}{\let\nl\oldnl}}
\usepackage[inline]{enumitem}

\usepackage[belowskip=-10pt,aboveskip=10pt]{caption}

\setlist[enumerate,1]{label=\textit{(\roman*)}}

\newcommand{\pktodo}[1]{\todo[inline, color=orange!40]{PK: #1}}

\renewcommand{\paragraph}[1]{\vspace{5pt}\noindent\textit{#1}}

\newcommand{\fault}{fault\xspace}
\newcommand{\Fault}{Fault\xspace}
\newcommand{\faults}{faults\xspace}

\newcommand{\faulty}{faulty\xspace}

\begin{document}

\copyrightyear{2018}
\acmYear{2018}
\setcopyright{rightsretained}
\acmConference[ICSE-NIER'18]{40th International Conference on Software Engineering: New Ideas and Emerging Results Track}{May 27-June 3, 2018}{Gothenburg, Sweden}
\acmDOI{10.1145/3183399.3183402}
\acmISBN{978-1-4503-5662-6/18/05}

\title{Learning Spreadsheet Smells for \Fault Prediction}

\title{Improving \Fault Prediction for Spreadsheets based on Smells with a Machine Learning Approach}
\title{Improving Smell-based \Fault Prediction for Spreadsheets}
\title{A Machine-Learning Approach to Combining Spreadsheet Smells for Improved \Fault Prediction}
\title{Combining Spreadsheet Smells for Improved \Fault Prediction}


\author{Patrick	Koch}
\author{Konstantin Schekotihin}
\author{Dietmar Jannach}

\affiliation{%
	\institution{AAU Klagenfurt, Austria}
}
\email{[firstname].[lastname]@aau.at}

\author{Birgit Hofer}
\author{Franz Wotawa}
\affiliation{%
	\institution{Graz University of Technology}
	\country{Austria}
}
\email{{bhofer,wotawa}@ist.tugraz.at}

\author{Thomas Schmitz}
\affiliation{%
	\institution{TU Dortmund}
	\country{Germany}
}
\email{thomas.schmitz@tu-dortmund.de}

\renewcommand{\shortauthors}{Koch et al.}

\keywords{Spreadsheet Smells, Spreadsheet QA, Fault Prediction} 

\begin{abstract}
Spreadsheets are commonly used in organizations as a programming tool for business-related calculations and decision making.
Since \faults in spreadsheets can have severe business impacts, a number of approaches from general software engineering have been applied to spreadsheets in recent years, among them the concept of code smells. Smells can in particular be used for the task of \fault prediction.
An analysis of existing spreadsheet smells, however, revealed that the predictive power of individual smells can be limited.
In this work we therefore propose a machine learning based approach which combines the predictions of individual smells by using an AdaBoost ensemble classifier.
Experiments on two public datasets containing real-world spreadsheet \faults show significant improvements in terms of fault prediction accuracy. 


\end{abstract}

\maketitle

\section{Introduction}
Many decisions in organizations are based on spreadsheets.
One reason for the broad success of spreadsheets is their simple and intuitive computation paradigm, which allows even end users to develop spreadsheet programs according to their needs.
However, these programs are particularly prone to faults for two main reasons:
\begin{enumerate*}
	\item most of the users have no or only little background in general software development, and

	
	\item today's spreadsheet environments have limited support for quality assurance (QA).
\end{enumerate*}
The resulting \faults can lead to substantial financial losses for companies.\footnote{See \url{http://www.eusprig.org/horror-stories.htm} for a list of examples.}

Various quality assurance approaches for spreadsheets were suggested in recent years, including techniques for visualization, testing, debugging, and fault prevention~\cite{JannachSchmitzEtAl2014}.
\emph{Spreadsheet smells} are a prominent approach that can be particularly helpful in the context of \fault prevention, e.g., in preventive maintenance or \fault prediction.
They transfer the idea of \emph{code smells}~\cite{Fowler} to the spreadsheet domain and represent heuristics that are designed to indicate potential problems in spreadsheets such as complex formulas, possibly missing inputs, and problematic dependencies~\cite{CunhaFRS12,HermansPD12a,HermansPD12b,icsm/AbreuCFMPS14}.

Abreu \textit{et~al.}~\cite{icsm/AbreuCFMPS14} relied on a combination of spreadsheet smells and other techniques for  \emph{fault prediction}.
In particular, they used smells to derive a \fault likelihood for each cell in a spreadsheet.
Our work continues this general line of research on fault prediction using smells.
While Abreu \textit{et~al.} considered individual smells as equal in terms of predictive power, our research indicates that
\begin{enumerate*}
	\item the fault prediction power varies significantly across different smells, and
	\item the predictive power of individual smells is comparably low.
\end{enumerate*}

In this work, we therefore propose a novel smell-based \fault prediction approach for spreadsheets that is based on learning optimal combinations of smells with machine learning (ML) techniques.
Technically, we frame the smell-based \fault prediction problem as a supervised classification problem.
The inputs to the ML problem are
\begin{enumerate*}
	\item a set of spreadsheets as training data for which the \faulty formulas are known and
	\item a set of smells from the literature as \fault predictors.
\end{enumerate*}
The overall process of making predictions then consists of the following main steps.
First, we compute the ``strength'' of each given smell for all formulas that are contained in the training spreadsheets.
These smell values, together with a label (correct or \faulty), for each formula are then used as training data for the learning problem.
Given that form of data representation, a variety of supervised ML algorithms can be applied to learn a function to predict the \fault probabilities of unlabelled formulas.

We tested our method on two publicly available datasets of real-world spreadsheets for which the \faulty formulas are known.
The experimental evaluation showed that an ensemble method, Ada\-Boost~\cite{FeundSA99}, led to the best classification results and outperformed \fault predictors that were based on individual smells by far. The obtained absolute \emph{recall} values ranged between 70\,\% and 95\,\%, which indicates that a large majority of the existing \faults can be identified by the smell-based ensemble predictor.

\section{Related Work}
The work of Abreu \textit{et~al.}~\cite{icsm/AbreuCFMPS14} is the contribution that is most closely related to ours.
As mentioned in the introduction, the authors use smells as part of their \fault prediction approach for spreadsheets.
Similar to our work, they rely on a set of smells from the literature and, in a first step, compute the strength of each smell for each cell in the given spreadsheet(s).
The subsequent steps are, however, different from our work.
Abreu \textit{et~al.} apply a threshold for each computed measure that classifies each cell as being smelly or not.
They then compute the set of output cells (cells that are not referred to), as well as the calculation chains of these cells.
In a final step, Spectrum-based Fault Localization (SFL) is used to compute the suspiciousness of each cell.
Cells that are often involved in calculation chains of smelly cells and less often in calculation chains of non-smelly cells are more suspicious of being faulty.


Singh \textit{et~al.}~\cite{singhLZ17} proposed an approach to use machine learning methods for \fault prediction for spreadsheets.
In their tool, named {Melford}, a neural network is trained with a set of custom engineered features, based on the structure and content of spreadsheets, in order to predict \enquote{number-where-formula-expected} faults.
Differently from their work, our approach
\begin{enumerate*}
	\item uses spreadsheet smells from the literature as features,
	\item applies a different learning model, and
	\item is not limited to certain types of \faults.
\end{enumerate*}

A number of previous works considered code smells as part of the fault prediction process in the general field of software engineering.
Fontana \textit{et~al.}~\cite{FontanaMZM16}, for example, applied ML algorithms to detect code smells in software systems.
Palomba \textit{et~al.}~\cite{PalombaZFDO16} improved the performance of a bug prediction system based on smells by introducing the concept of ``smell intensity levels''.
Ma \textit{et~al.}~\cite{MaCZX16} used fault prediction based on smells to guide the refactoring of code.
While these approaches aim to improve prevalent software QA practices, they were not designed to consider the specific types of potential problems that can be found in spreadsheet programs.
\section{Technical Approach}\label{sec:learning}
In this section, we provide the technical details of how we framed the fault prediction problem as a \emph{supervised classification} problem using spreadsheet smells, how we preprocessed the data, and how we optimized the used prediction models.

\paragraph{Problem Definition and Data Preprocessing.}
The fault prediction problem can be summarized as follows: Given
\begin{enumerate*}
	\item a set of \faulty spreadsheets in which every formula is labeled either as being \faulty or correct, and
	\item a number of smells,
\end{enumerate*}
learn a function that predicts whether some previously unseen formula is \faulty or not.

Supervised ML techniques use a set of training examples, where each example is characterized by a set of features and has one label assigned.
In our case, each example is constructed for a formula of a training spreadsheet and its label indicates the formula being \faulty or correct.
The set of features corresponds to the set of smells that are used in the learning problem.
The feature values (called the feature vector) for each example are determined by computing the strength of each smell for the related formula.
Table~\ref{tab:learning_data} shows the general structure of the problem encoding.

To build the table of training data, for each formula we compute the strength of each smell according to the heuristics from the literature, and assign the appropriate label provided in the input.
The resulting training data table is complete, i.e., no value is missing.

\paragraph{Model Optimization \& Learning.}
Given these inputs, a variety of machine learning approaches can be applied, optimizing some given performance measure.
Since the given classification problem is binary (\faulty or correct), we optimize our models for the F1-measure, which is a standard classification accuracy measure that is computed as the harmonic mean of \emph{precision} and \emph{recall}.

\begin{table}[t]

\caption{Structure of the training data}
\label{tab:learning_data}
\centering
\begin{tabular}{l|c|c|c|c} \hline
Cell & $Smell_1$ & $\dots$ & $Smell_n$ & $Label$ \\ \hline
$cell_1$  & $value_{1,1}$ & $\dots$ & $value_{n,1}$ & correct/faulty\\
$\dots$  & $\dots$ & $\dots$ & $\dots$ & $\dots$\\
$cell_n$  & $value_{1,n}$ & $\dots$ & $value_{n,n}$ & correct/faulty \\ \hline
\end{tabular}

\end{table}

We tested various ML techniques for the given problem.
The best results in terms of F1-measure and high recall were achieved when we used \emph{Adaptive Boosting} (AdaBoost)~\cite{FeundSA99}, and we therefore use it as representative in our evaluation.
AdaBoost is a meta-algorithm that combines the output of many, possibly individually weak, classifiers (in our case decision trees) to obtain a better classification outcome.


Supervised learning techniques allow for the fine-tuning of an optimization goal for the given data using model-specific parameters.
In our experiments, we apply a grid search method to explore all possible values from a given set and pick the one that leads to the highest value of the F1-measure.
In the case of AdaBoost, the main parameter to be set was the number of used decision trees.

We use 10-fold cross-validation for optimizing and evaluating our models.
To avoid that results are dependent on the choice of the partitioning into training and test examples, we apply stratified folding with shuffling, guaranteeing a mixed but roughly equal distribution of samples of both classes within each fold.
Before processing, the feature values are standardized, shifting the data for each feature to zero mean and scaling it to Gaussian unit variance in order to meet the requirements of the used supervised learning approaches.
Finally, since the number of correct formulas in the input spreadsheets is significantly higher than the number of \faulty ones, models trained with this data might be biased to predict an input formula to be correct.
Therefore, we use an \emph{oversampling} procedure which is applied to the training data of every fold.
Specifically, we generate additional training examples from the minority class, i.e., cases that are labeled as \faulty, by adding copies of randomly picked \faulty examples until the number of examples in each class is equal.

\section{Evaluation}
We performed experiments on two datasets to assess the effectiveness of our smell-based \fault prediction method.
We recorded the F1-measure using a 
10-fold cross-validation procedure, 
and compare our performance results using AdaBoost with those that were achieved when individual smells were used either as predictors or as part of a voting committee, and with the results of using an alternative machine learning method.
To enable validation and replicability of our research, we share the source code used in the experiments and the detailed results for all datasets online\footnote{\url{http://spreadsheets.ist.tugraz.at/wp-content/uploads/2018/01/ICSE18.zip}}.

\subsection{Study Setup}
\paragraph{Datasets.} The first dataset is based on a subset of the Enron spreadsheet corpus~\cite{icse/HermansM15} which contains real-world \faults~\cite{SchmitzJannach2016}; the detailed list of \faults can be found online.\footnote{\url{http://ls13-www.cs.tu-dortmund.de/homepage/spreadsheets/enron-errors.htm}}
Overall, the Enron Errors Corpus contains 26 spreadsheets with \faulty formulas, with 2.9\,\% of the formulas -- 481 out of 16,790 -- being faulty.
The second dataset (called ``INFO1'')~\cite{GetznerHW17} contains spreadsheets developed by civil engineering students as part of an exercise.
It comprises 119 spreadsheets with 5,157 \faulty formulas (3.0\,\%).
More details can be found online.\footnote{\url{http://spreadsheets.ist.tugraz.at/index.php/corpora-for-benchmarking/info1/}}

\paragraph{Used Smells.}
We used a set of 19 spreadsheet smells and the corresponding strength calculation rules that were proposed in previous research \cite{CunhaFRS12,HermansPD12a,HermansPD12b,icsm/AbreuCFMPS14}.
In general, the strength of a specific smell is expressed by the value of a related complexity metric that is measured for the given formula or worksheet.
The detailed list of used smells is shown in Table~\ref{tab:smells}.
Since we focus on the prediction of \faulty formulas, we included formula and worksheet smells and did not consider smells for data cells.
The measurements of the worksheet smells were applied to each formula of the worksheet.
\begin{table}[ht]
	\centering
	\caption{Overview of Used Smells}
	\label{tab:smells}
	\rowcolors{2}{gray!15}{white}
	\scalebox{0.9}{
	\begin{tabular}{llc}
		\toprule
		Index & Name & Target\\
		\midrule
		0  & Column-wise Pattern Finder \cite{CunhaFRS12} & cell\\
		1  & Row-wise Pattern Finder \cite{CunhaFRS12} & cell\\
		2  & Reference to empty cells \cite{CunhaFRS12} & cell\\
		3  & Changing Formulas \cite{HermansPD12a} & cell\\
		4  & Changing Worksheets \cite{HermansPD12a} & cell\\
		5  & Duplicated Calculations \cite{icsm/AbreuCFMPS14} & cell\\
		6  & Duplicated Formulas \cite{HermansPD12b} & cell\\
		7  & Feature Envy \cite{HermansPD12a} & cell\\
		8  & Long Calculation Chain \cite{HermansPD12b} & cell\\
		9  & Conditional Complexity \cite{HermansPD12b} & cell\\
		10  & Multiple Operations \cite{HermansPD12b} & cell\\
		11  & Multiple References \cite{HermansPD12b} & cell\\
		12  & Inappropriate Intimacy \cite{HermansPD12a} & worksheet\\
		13  & Middle Man \cite{HermansPD12a} & worksheet\\
		14  & Shotgun Surgery (Formulas) \cite{HermansPD12a} & worksheet\\
		15  & Shotgun Surgery (Worksheets) \cite{HermansPD12a} & worksheet\\
		16  & Inconsistent Formula Group Reference\cite{Koch2016} & worksheet\\
		17  & Missing Header\cite{Koch2016} & worksheet\\
		18  & Overburdened Worksheet\cite{Koch2016} & worksheet\\
		
		\bottomrule
	\end{tabular}
	}
\end{table}


\paragraph{Baseline Methods.}
To assess the performance of the AdaBoost classifier, we compare it with three types of baselines.
The first type uses individual smells and implements a simple classification rule.
Given a formula, a spreadsheet smell, and a threshold percentage $T$, it classifies the formula as \faulty if the computed strength of the smell lies above the lowest $T$\,\% of all feature values for the smell.
We determined the optimal value for $T$ for each of the 19 smells through a grid search method.
Following the suggestion of Hermans \textit{et~al.}~\cite{HermansPD12a}, we tested three threshold values for $T$ (70\,\%, 80\,\%, and 90\,\%).
For the second baseline, we combined the optimized predictions of individual smells using two simple voting schemes:
\begin{enumerate*}
	\item majority voting using uniform weights (called \enquote{Voting: majority}), and
	\item advocate voting, where any smell classifier voting for \faulty suffices for the ensemble to classify a sample as \faulty (called \enquote{Voting: advocate}). 
\end{enumerate*}
As the third baseline, we use linear Support Vector Machines (SVMs), as they were found to be effective for comparable learning tasks \cite{FontanaMZM16,Maiga2012}.
To deal with the computational complexity in particular for the larger INFO1 dataset, we chose Stochastic Gradient Descent (SGD) as the learning method for the used linear SVMs, which is recommended for large-scale training of classifiers \cite{bottou2010large}.
We optimized the regularization parameter $\alpha$ of the SGD training process through a systematic grid search.
The first two baselines model a scenario in which a user manually selects one smell or a combination of smells for fault detection.
The third baseline offers a comparison with another established ML approach.


\paragraph{Parameter Selection.}
For the AdaBoost classifier, the grid search using a 10-fold cross-validation procedure returned 5 as the optimal number of decision trees for the Enron Errors Corpus and 1 for the INFO1 corpus. 
The optimized $T$ values for the individual smell classifiers can be inspected in the analysis script provided online.
The \enquote{voting} classifiers  use the already optimized classifiers of individual smells. 
For the SVM baseline with SGD, using a regulatory parameter of 0.0001 led to the best results for the Enron dataset; 0.001 was the optimal setting for the INFO1 dataset.

\subsection{Results and Discussion}
\begin{figure}[bht]
	\centering
	\includegraphics[trim=1.8cm 0cm 0cm 0cm, width=1.0\columnwidth]{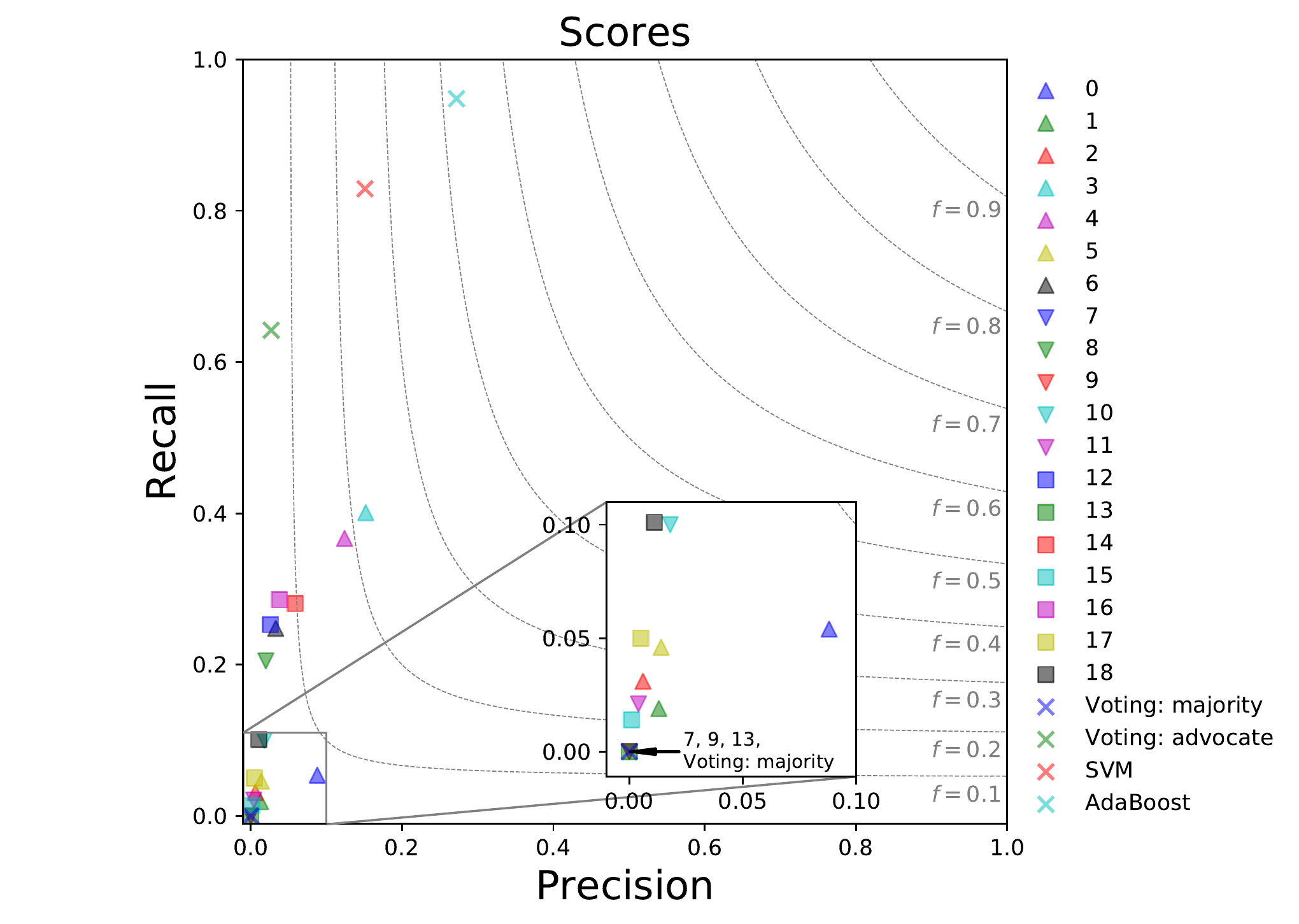}
	\caption{Precision-recall performance for the Enron Errors Corpus. The numbers in the legend correspond to the indices given in Table~\ref{tab:smells}.}
	\label{fig:scores_enron}
\end{figure}

Figure~\ref{fig:scores_enron} shows the results for precision (x-axis), recall (y-axis) and the F1-measure (radial line) obtained for the Enron Errors Corpus.
Smells that target cells are represented by triangles, worksheet-based smells are represented as squares, and the results of voting and ensemble classifiers are indicated by \enquote*{x} symbols.

The proposed ensemble learning approach, AdaBoost, significantly outperforms the baseline techniques. 
The obtained recall value is at about 95\,\%,
which means that the majority of \faults was successfully identified by our method.
The precision of about 30\,\% implies that two out of three \fault predictions are \enquote{false alarms}.
Whether precision or recall is more important depends on the domain or application scenario.
In our case, the main goal is not to miss \faults that otherwise would remain in the spreadsheets. 
Hence, high recall values are particularly desirable. 
While a \enquote{false alarm} rate of 70\,\%  might seem a lot, consider that each examined spreadsheet contains usually only one to three \faults.
Hence, only about 3 to 9 of possibly hundreds of formulas have to be inspected.
Moreover, these results were achieved using a fixed set of smell-based predictors and a limited set of training data.
Diversification and optimization of the used features, as well as the use of additional training data might further improve the precision scores.

While outperformed by AdaBoost, SVM performs well, achieving a recall of about 83\,\%, and a precision of about 15\,\%.
This confirms the conjecture that more elaborate ensemble methods generally perform better than any single classifier~\cite{Dietterich00}.

In comparison, the simple combination of smell classifiers by means of voting schemes, as indicated by the results of the \enquote{voting} classifiers, lead to poor prediction performance.
Majority voting did not detect any faults, as no majority was found for any of the faulty cells.
Advocate voting achieved a recall of about 65\,\%, but only a precision of about 3\,\%.
This reveals the major shortcomings of simple ensemble schemes using smells: no single threshold-based smell classifier is capable of detecting all faults, and no fault case is pronounced enough for a majority of smells to indicate it.

Many of the individual smells 
have limited predictive power when used in isolation, leading to low recall and precision values.
Smells that are measured per formula cell, barring some exceptions, generally exhibit limited prediction performance.
Smells that are measured per worksheet slightly outperform the majority of per-formula smells in terms of recall, but also lack precision.

Overall, the use of isolated smells and their simple combinations is not very helpful for \fault prediction, whereas combining them as proposed in this work leads to substantially higher predictive power.
This indicates that actual \faults in spreadsheets emerge from a combination of specific deficiencies which are difficult to capture by means of simple metric thresholds.

The evaluation on the INFO1 dataset led to comparable results\footnote{A precision-recall plot can be found in the online material.}:
The best performing classifier is AdaBoost (recall: 71\,\%, precision: 30\,\%, F1: 0.42).
While SVM and the advocate voting ensemble have a higher recall (77\,\% respectively 78\,\%),
their precision is significantly lower (7\,\% and 3\,\%), as is their F1 score (0.12 and 0.06).
The majority voting ensemble has both a precision and a recall of 0\,\%.
The F1-measure of all individual smell classifiers is below 0.1.
These results confirm the ones we have obtained for the Enron Errors Corpus.

\subsection{Threats to Validity}
The main threat to the \emph{internal} validity of our research is related to the correctness of the software used for analysis and evaluation.
To allow other researchers to validate our work, all source code and the used datasets are provided online.
The main threat to the \emph{external} validity of our study is the representativeness of the used spreadsheet corpora with regard to the overall population of \faulty spreadsheets.
Generally, the Enron spreadsheets used in the study have been used extensively for empirical research in previous works.
The specific set of real-world \faults in the corpus was furthermore obtained in a systematic and reproducible manner \cite{SchmitzJannach2016} and we therefore consider the risk that the \faults are not representative as low.
The representativeness of the INFO1 corpus might be limited as it contains spreadsheets that were designed
for the same problem specification.
Nonetheless, since the obtained results are very similar for both datasets, we are confident that the observations obtained with this dataset are reliable as well.


\section{Conclusions \& Outlook}\label{sec:concl}
Our work shows that spreadsheet smells can be valuable instruments for \fault prediction in spreadsheets when they are not considered in isolation.
In general, we consider the application and further development of modern and powerful machine learning methods for spreadsheet quality assurance as an emerging and promising area, in particular as past approaches to spreadsheet QA were often based on heuristics for \fault identification and repair that were designed based on domain expertise.

From an algorithmic perspective, our next steps include the investigation of alternative learning models, in particular deep-learning techniques, the application of feature selection methods to identify and remove noisy smells, and the exploration of alternative methods for oversampling. Regarding the general approach, we plan to investigate the performance of additional types of smells and other spreadsheet quality metrics as \fault predictors.

\begin{acks}
The work described in this paper has been been funded by the Austrian Science Fund (FWF) project {\em DEbugging Of Spreadsheet programs (DEOS)} under contract number I2144 and the Deutsche Forschungsgemeinschaft (DFG) under contract number JA 2095/4-1.
\end{acks}

\bibliographystyle{ACM-Reference-Format}
\bibliography{main}

\end{document}